\documentclass[aps,reprint,superscriptaddress,showkeys,floatfix]{revtex4-2}
\usepackage[T1]{fontenc}
\usepackage[utf8]{inputenc}
\usepackage{amsmath,amssymb,upgreek}
\usepackage{graphicx}
\usepackage{color}
\newcommand*{\etal}{\textit{et al.}}
\newcommand*{\rpm}{\,\mathrm{rpm}}

\newcommand*{\cm}{\,\mathrm{cm}}
\newcommand*{\wtpc}{\,\mathrm{wt.\%}}
\newcommand*{\mPas}{\,\mathrm{mPa\,s}}
\newcommand*{\micron}{\,\upmu\mathrm{m}}
\newcommand*{\Reynolds}{\mathrm{Re}}
\newlength\figheight
\begin{document}
\title{Droplet size distribution in emulsions}

\author{Manon L'Estim\'e}
\email{Corresponding author: manonlestime@gmail.com}
\affiliation{Van der Waals-Zeeman Institute, Institute of Physics, University of Amsterdam, 1098XH Amsterdam, The Netherlands.}
\author{Michael Schindler}
\affiliation{CNRS~UMR7083, ESPCI~Paris, Universit\'e~PSL, 10~rue~Vauquelin, 75005 Paris, France}
\author{Noushine Shahidzadeh}
\affiliation{Van der Waals-Zeeman Institute, Institute of Physics, University of Amsterdam, 1098XH Amsterdam, The Netherlands.}
\author{Daniel~Bonn}
\affiliation{Van der Waals-Zeeman Institute, Institute of Physics, University of Amsterdam, 1098XH Amsterdam, The Netherlands.}

\date{\today}

\begin{abstract}
\noindent \textbf{Abstract:} The droplet size in emulsions is known to affect the rheological properties and
plays a crucial role in the many applications of emulsions. Despite its
importance, the underlying mechanisms governing droplet size in emulsification
remain poorly understood. We investigate the average drop size and size
distribution upon emulsification with a high-shear mixer for model oil-in-water
emulsions stabilized by a surfactant. The size distribution is found to be a
log-normal distribution, resulting from the repetitive random breakup of drops.
High-shear emulsification, the usual way of making emulsions, is therefore
found to be very different from turbulent emulsification given by the
Kolmogorov--Hinze theory for which power-law distributions of the drop size are
expected.  In agreement with this, the mean droplet size does not follow a scaling with the Reynolds number of the emulsification flow, but rather a capillary number scaling based on the viscosity of the continuous phase.
\end{abstract}

\maketitle
\section*{Introduction}

An emulsion is a dispersion of droplets in a continuous phase produced by
dispersing one fluid into another immiscible fluid. The rheological properties
of emulsions are of great interest for instance for food and cosmetic products
\cite{deblais2021, brummer1999}, above all for the homogeneity of the flow of
such materials \cite{PhysRevLett.105.225502}. For concentrated emulsions, the
rheology is determined by the Laplace pressure~$\gamma/r$ that indicates the
deformability of individual droplets; here $\gamma$~is the interfacial tension
and $r$~is the droplet radius \cite{princen1986rheology, pal1996effect,
dekker2018}. Consequently, the rheology can be tuned by changing the size of
the droplets which is a key factor for the quality of the final product as it
can affect its stability, flavor, texture and mouthfeel~\cite{chung2014}.

Since the pioneering work of Hinze and Kolmogorov on the breakup of droplets in
a turbulent flow \cite{hinze1955fundamentals, kolmogorov1949}, numerous studies
have been reported on the droplet size distribution in emulsions. A large part
of the work focuses on the impact of the formulation variables, such as the
viscosities of the fluids \cite{calabrese1986drop, Rodgers2017effect,
rueger2013dispersion}, the volume fraction of the dispersed phase
\cite{hall2011droplet} or the interfacial tension \cite{walstra1993principles,
soligo2019breakage}. It was shown recently that the rheology of simple
emulsions can be understood on the basis of the volume fraction, interfacial
tension and drop radii, so that knowing the drop radii allows to predict the
rheology~\cite{maureen} if the interfacial tension is known. We therefore need
to establish what governs the drop size.

Various mechanistic models have been proposed to characterize and predict the
droplet size distribution. Most of these rely on the Kolmogorov--Hinze theory
of turbulent emulsification in which the breakup mechanisms depend on the
Kolmogorov length scale given by the size of the smallest eddies. Droplets
larger than this length will break up under the action of turbulent inertial
stress induced by the pressure fluctuations along the drop surface, smaller
ones remain intact. At scales smaller than the Kolmogorov length scale,
cohesive forces resulting from interfacial tension and drop viscosity oppose
drop fragmentation. Hence, the maximal stable droplet diameter results from the
equilibrium between the internal cohesive and the external turbulent stresses,
and the drop size therefore depends on the Reynolds number (or energy dissipation
rate). In simulations, turbulent emulsification has been studied in detail
recently \cite{yi_toschi_sun_2021, Luca}, for which power-law distributions of
the drop size were found in agreement with the idea that the turbulent energy
cascade is important for drop formation \cite{Luca}. However also log-normal
and Gamma function drop size distributions were reported
\cite{yi_toschi_sun_2021}, so that it is unclear what determines the size
distribution. In an extensive study, Vankova et al.~\cite{VANKOVA2007363}
showed that although the average drop size is well described by Kolmogorov--Hinze theory,
the drop size distribution is well fitted by a log-normal distribution that
does not follow from any turbulent emulsification theory. In addition, while
this theory identifies a clear emulsification mechanism, it assumes a
homogeneous and isotropic turbulence that is hardly achieved during the process
\cite{haakansson2019}. The presence of multiple phases will affect the
turbulence itself, as known from turbulent drag reduction
\cite{PhysRevE.47.R28}. In dense emulsions it is impossible to
separate the flow of the continuous phase from the motion and deformation of
the discrete phase.

An alternative to the turbulent emulsification theories are fragmentation
theories, that describe the breakup due to either surface tension or the drag
with the continuous phase, without necessarily requiring a turbulent
flow. There are two competing fragmentation theories for the droplet size
distribution. First, the breakup of droplets can be viewed as a sequence of
random multiplicative processes resulting in a drop size distribution that is
well described by a log-normal distribution \cite{kolmogorov1941,
villermaux2007}:
\begin{equation}
  \label{eq:lognormal}
  \mathcal{P}\Bigl(x{=}\frac{R}{\langle R\rangle}; \mu,\sigma\Bigr)
    = \frac{1}{x \sigma \sqrt{2 \pi}}
      \exp \left( \frac{-(\ln{x} - \mu)^2}{2 \sigma^2} \right).
\end{equation}

Second, one may have liquid threads (``ligaments'') forming through the
Kelvin--Helmholtz instability, that subsequently break up into droplets due to
the surface tension. As demonstrated by Villermaux
\etal~\cite{villermaux2004ligament}, the breakup of ligaments leads to a Gamma
distribution of sizes, namely
\begin{equation}
  \label{eq:gamma-n}
  \mathcal{P}\Bigl(x{=}\frac{R}{\langle R\rangle}; n\Bigr)
    = \frac{n^n}{\Gamma(n)}x^{n-1}e^{-nx},
\end{equation}
where $\Gamma$~is the gamma function and $n$~is a parameter that depends on the
ligament corrugation.

Also the contrary of fragmentation is possible, coalescence of small
droplets into large ones. In principle, both processes may happen, but their
rates are very different. The main reason for this asymmetry is the presence of
surfactants, hindering coalescence. As a more subtle mechanism, we have no
reason to assume these rates are constants. If they depend on the current state
of the flow itself, and thus on the droplet size distribution, then the
fragmentation follows a non-linear process, allowing the dominance of
fragmentation over coalescence.

In practical situations mostly rotating emulsifiers such as the one used in the
present study are used. For such setups, the influence of several parameters
have been investigated: Speed and type of impeller were looked at, as well as
the location of the dispersed phase addition \cite{Rodgers2017effect,
el2009dispersion}. In addition, different emulsification devices were compared
\cite{james2017scale}, as well as the mode of operation (batch or continuous)
\cite{haakansson2018rotor}. However, no clear unified picture of the drop size
and its distribution emerged from these studies.

In the present work we investigate how the droplet size, its mean and its
distribution are both influenced by the impeller speed, by the system
formulation, and by the mixer geometry. This allows us to  propose a simple
scaling for the mean droplet size that does not invoke the turbulent energy
cascade, and allows us to distinguish between the two fragmentation models,
favoring the random breakup scenario.

\section*{Materials and methods}
\subsection*{Preparation of the emulsions}
\begin{figure}[t]
    \centering
    \includegraphics[height=\figheight]{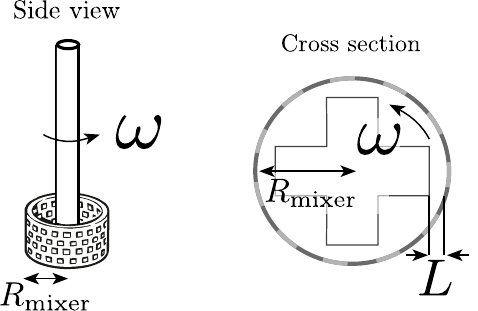}
    \caption{Main components of the mixer. The cross-shaped impeller rotates at
    a speed~$\omega$ and is surrounded by the stator. The latter is an open
    cylinder of radius~$R_\mathrm{mixer}$ whose surface is covered by small
    holes. The distance between the blades and the cylinder is denoted~$L$. The
    latter differs slightly according to the inner radius of the cylinder:
    $L=0.25$ or $0.3\:\rm{mm}$ when $R_\mathrm{mixer}=0.8$ or
    $1.6\:\rm{cm}$, respectively.}
    \label{fig1}
\end{figure}

Oil-in-water emulsions are prepared using a Silverson high shear rotor/stator
laboratory mixer~(LM5-A). As  depicted in Fig.~\ref{fig1}, the mixer
is composed of a rotor with a cross-shaped impeller spinning at a
speed~$\omega$. The impeller is surrounded by the stator, an open cylinder with
a surface covered by small squared holes. We denote by $R_\mathrm{mixer}$~the
inner radius of the stator and by $L$~the distance between the rotor
extremities and the inner part of the cylinder.
\begin{figure*}[t]
    \includegraphics[height=5cm]{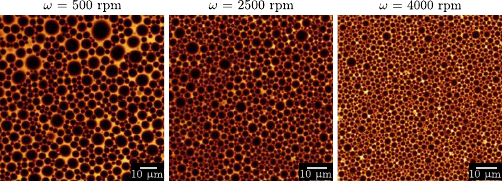}
    \caption{Confocal images of an emulsion at successive rotation speeds
    $\omega = 500$, $2500$ and $4000\rpm$. The emulsion has been prepared
    using a mixer of radius $R_\mathrm{mixer}=1.6\:\rm{cm}$ to disperse $80$\%
    of castor oil in an aqueous solution of viscosity $\eta_c=1\mPas$,
    containing $3\wtpc$ of~SDS. The oil droplets appear in black by
    contrast with the continuous phase that contains Rhodamine~B dye. }
    \label{fig2}
\end{figure*}%

We use castor-oil-in-water emulsions stabilized by sodium dodecyl sulphate
(SDS) surfactant as our model emulsion. The dispersed phase consist of castor
oil of viscosity $\eta_d = 580\mPas$. The continuous phase is
prepared by dissolving $3\wtpc$ of SDS~surfactant in demineralized water.
Rhodamine~B is then added to the solution as a dye. As discussed
later, the large concentration of surfactants inhibits droplet coalescence and
ensures that the total amount of surfactant molecules present is larger than
the quantity needed to stabilize the emulsion.

To prepare the emulsions, we first gradually add oil to the aqueous solution
while stirring at $500\rpm$. Once we have reached the desired volume ratio, the
rotation speed is increased step-by-step up to $4000\rpm$. To homogenize the
mixing, the beaker is constantly rotated around the axis of the rotor.

\subsection*{Data acquisition}

During the process, for each rotation speed, samples of the emulsion are
collected and visualized using confocal laser scanning microscopy. These
pictures show circular sections through individual droplets, see
Fig.~\ref{fig2}. They are then detected and measured by the \emph{ImageJ}
software with its plug-in \textit{``analyse particle''}~\cite{ImageJ,AnalyseParticle}.

\subsection*{Data treatment}

From the observed 2D~circle radii we want to conclude on the distribution of
3D~spherical radii of the droplets. This challenge has several aspects, namely
(i)~the question whether and how it is possible to obtain the original
3D~distribution from a distribution of 2D~sections -- assuming that we deal
with nice smooth distribution densities. Then, (ii)~the question how to repeat
the same task on a finite number of observed 2D~radii.

In the following, $R$~denotes 3D~radii of spheres, distributed according to a
probability density~$\rho_R$. The radii and their distribution of 2D~sections
are denoted $r$ and~$\rho_r$, respectively.

In order to answer question~(i), we assume that when an emulsion of spheres is
cut by a confocal plane, any sphere of radius~$R$ is cut \emph{statistically
independently} uniformly at height~$z$ in the interval~$[-R,R]$. This implies
some independence in the positions of the spheres which are rather implicit and
are beyond the scope of the present work. The combined probability density to
find a sphere of radius~$R$ cut at height~$z$ is then $\rho_c(R,z) = \rho_R(R)
\rho(z|R)$, with
\begin{equation*}
   \rho(z|R) = \left\{\begin{array}{cl}1/(2R) & \text{if $-R<z<R$}\\ 0 & \text{else}\end{array}\right\}.
\end{equation*}
The probability density~$\rho_r$ of the observed cut~radii~$r$ follows as
\begin{align*}
   \rho_r(r)
     &= \int_0^\infty dR \int_{-R}^{R} dz\: \rho_c(R,z) \delta\bigl(r - \sqrt{R^2 - z^2}\bigr) \\
     &= r \int_r^\infty dR \frac{\rho_R(R)}{R\sqrt{R^2 - r^2}}.
\end{align*}
We need to invert this integral equation and obtain $\rho_R$ from~$\rho_r$. To
ease this task, we work with squared radii. The corresponding integral equation
for the densities $\rho_{r^2}$ and $\rho_{R^2}$ is
\begin{equation*}
   \rho_{r^2}(x) = \frac{1}{2} \int_x^\infty dt \frac{1}{\sqrt{t-x}} \frac{\rho_{R^2}(t)}{\sqrt{t}}.
\end{equation*}
Up to the integral bounds, the inversion of this is known as \emph{Abel's
problem}, and a recipe how to solve it is given in Section~1.3.4 of
Ref.~\cite{PorSti90}. The result is
\begin{equation}
  \label{eq:inverted_sq}
  \rho_{R^2}(x) = -\frac{2}{\pi} \sqrt{x} \frac{d}{dx} \int_x^\infty dt \frac{\rho_{r^2}(t)}{\sqrt{t-x}}.
\end{equation}

A critical comment on this inversion is in order: We do not fully understand in
what function spaces it works. In particular, equation~\eqref{eq:inverted_sq}
does not guarantee~$\rho_{R^2}$ to be a distribution density: it can be
negative. On the other hand, it respected the total integral very well in all
cases we applied it to. We tested the method on artificially generated data,
and found that it works. In the measured data we found indeed negative values
in $\rho_{R^2}$. This problem was more pronounced in those datasets that do not
contain very small radii.

For the numerical treatment, and to answer the above question~(ii), we used the
best resolution on the input data, that is $\rho_{r^2}$ being the sum of delta
functions, representing the individual measured radii. The integral is then a
highly irregular function of~$x$, consisting of many (weak) singularities~$\sim
1/\sqrt{\rule{0.5em}{0pt}}$. Before taking the derivative, we smoothed this
function with a Gaussian convolution kernel; its width was chosen such that the
result shows well the global behaviour of the curve, without too much
fluctuations.

When applied to the mean radii, the above equations predict that
\begin{equation}
  \langle r\rangle = \frac{\pi}{4}\langle R\rangle.
  \label{eqn:eq4}
\end{equation}

We tested this relation in a preliminary experiment in which we analyzed confocal cuts of a transparent emulsion at successive heights. We used a silicone oil-in-glycerol/water emulsion stabilized by SDS~surfactant. The
continuous phase is prepared by dissolving $3\wtpc$ of SDS in a 50:50 mixture
of glycerol and demineralized water. The dispersed phase consist of silicone
oil of viscosity $\eta = 500\mPas$, in which Nile red is added as a dye. Owing
to the addition of glycerol to the water, the refractive indexes of the two
phases are matched thus the emulsion is transparent. We proceed as previously
to prepare an emulsion composed of $80\wtpc$ of oil. Samples are collected at
rotation speeds $\omega = 3000$, $5000$ and $7000\rpm$, and analyzed by
confocal microscopy to measure the droplets radii at successive heights.

\begin{figure}[h!]
    \centering
    \includegraphics[height=5cm]{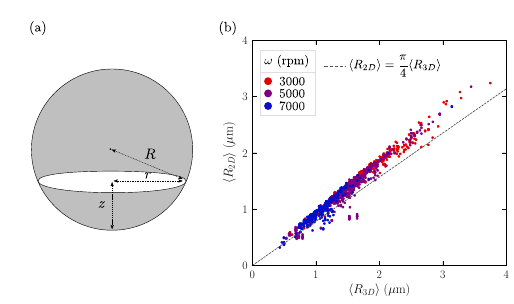}
    \caption{(a) The radius $r$ of a sphere's cross section at a random height
    $z$ is likely to be smaller than the true radius $R$ of the sphere. (b)
    Experimental determination of the correction factor by averaging the
    droplet's 2D radii measured on a cross-section as a function of the average
    of their 3D radii, using confocal images of a transparent emulsion at
    different heights. The data points include three samples corresponding to
    rotation speeds $\omega$ of $3000$ (red dots), $5000$ (purple dots), and
    $7000\rpm$ (blue dots), where each dot corresponds to a cross-section. The
    dashed line of slope~$0.79$ corresponds to the theoretical correction
    factor derived in eq.\eqref{eqn:eq4}.}
    \label{fig:AppA}
\end{figure}

Each droplet is characterized by a set of 2D radii in which the larger one
approximates the ``true'' 3D~droplet radius. Fig.~\ref{fig:AppA} compares the
average of the observed 2D radii, $\langle R_{2D} \rangle$, with the average of
the corresponding 3D radii, $\langle R_{3D} \rangle$, for each cross section.
The graph includes data from the three sample, each being characterized by a
rotation speed $\omega = 3000$ (red dots), $5000$ (purples dots) or $7000\rpm$
(blue dots). The dashed line of slope~$0.79$ corresponds to the theoretical
correction factor.

The data collapse on line of slope 0.91, larger than the slope predicted by the theory. This discrepancy is most likely due to the estimation of the 3D droplet radius, as the larger value of the set of 2D radii tends to slightly underestimate the true radius of the droplet. 
Nevertheless, the linear relationship between $\langle R_{2D} \rangle$ and $\langle R_{3D} \rangle$ still holds.

\section*{Results and Discussion}
In a first experiment, we demonstrate that the coalescence effects are hindered by the large concentration of surfactants. Thereafter, we vary the rotation speed~$\omega$, the oil volume fraction~$\phi$, the mixer geometry and the viscosity~$\eta_c$ of the continuous phase to find the key ingredients determining the droplet size.
\begin{figure}[h!]
  \centering
  \includegraphics[height=6cm]{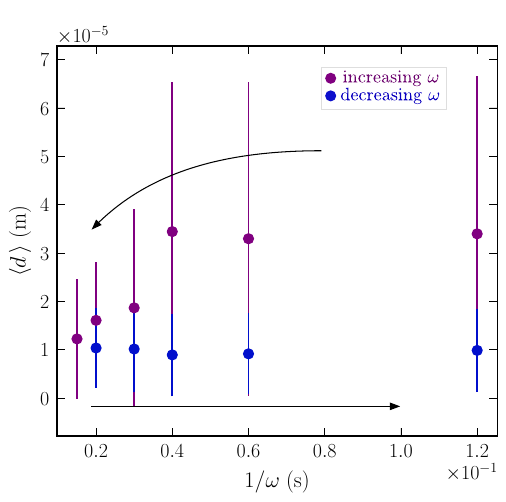}
  \caption{Demonstration of the absence of coalescence. The rotation speed
  $\omega$ is changed according to the arrow directions, leading to a
  hysteresis cycle of the mean droplet diameter~$\langle d\rangle$. The
  emulsion contains $3\wtpc$ of SDS and $40\wtpc$ of Castor
  oil, it has been prepared with a mixer of radius $R_\mathrm{mixer}=0.8\:\text{cm}$
and a continuous phase of viscosity $\eta_c=1\mPas$. The vertical bars report the standard deviation.}\label{fig:hysteresis}
\end{figure}
\subsection*{Ensuring a negligible droplet coalescence}%

The high concentration of surfactants in the system is expected to prevent
droplet coalescence. In order to confirm this, we conducted an experiment in
which the rotation speed was first increased and then decreased in a stepwise
manner, while samples were collected. Figure~\ref{fig:hysteresis} shows the variation
of the mean droplet diameter~$\langle d\rangle$ with the rotation
speed~$\omega$. The (upper) purple dots correspond to increasing speed (from right to
left) and the (lower) blue dots correspond to decreasing speed (from left to right).
The arrows indicate the succession of the data points.
Initially, the increase of the impeller speed leads to a reduction in droplet
size. However, during the subsequent decrease of the rotation speed, this
change is not reversed, but instead the droplets remains small. We interpret
this as a result of the high concentration of surfactants effectively
preventing droplet coalescence.

\subsection*{Influence of the rotation speed}

Fig.~\ref{fig2} shows three confocal images of the same emulsion at rotation
speeds $\omega=500$, $2500$ and $4000\rpm$. The dispersed oil droplets contain
Rhodamine~B dye and appear in black. The other parameters of the emulsion are
$R_\mathrm{mixer}=1.6\:\text{cm}$, $\phi=0.8$, $\eta_c=1\mPas$.
\begin{figure}
  \centering
  \includegraphics[height=6cm]{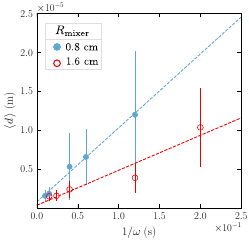}%
  \caption{Mean droplet diameter~$\langle d \rangle$ as a function of the
  inverse rotation speed~$1/\omega$ for emulsions prepared with
  $\phi= 0.8$, $\eta_c=1\mPas$ and $R_\mathrm{mixer} = 0.8\cm$ (blue dots) or
  $1.6\cm$ (red circles). The mean droplet diameter (dots) decreases with the mixer
  radius and scales as the inverse of the rotation speed. The standard
  deviation of the diameter (vertical lines) decreases likewise.}%
  \label{fig3}
\end{figure}
\begin{figure}
  \centering
  \includegraphics[height=6cm]{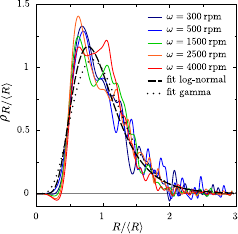}%
  \caption{The probability densities of the droplet sizes, as obtained from the
  inversion equation~\eqref{eq:inverted_sq}, for various rotation speeds
  $\omega$. The data are the same as the ones for $R_\mathrm{mixer}=1.6\cm$ in
  Fig.~\ref{fig3}. While the mean droplet diameter varies with the rotation
  speed, the shapes of the distributions are similar. The log-normal fit has
  $\mu=-0.0957$ and $\sigma=0.408$, and the gamma fit has $n=6.138$.}%
  \label{fig:inverted}%
\end{figure}

With increasing rotation speed also the number of droplets increases, while
their size becomes smaller. The mean droplet diameter~$\langle R \rangle$ is
calculated for each picture and plotted against the inverse rotation speed in
Fig.\;\ref{fig3} (red circles). The mean droplet diameter scales as $1/\omega$
and decreases from $10$ to $1\micron$ between $500$ and $4000\rpm$.
The vertical bars indicate the standard deviation which is rather
large especially at low rotation speed. Indeed, the complexity of such dense
systems inevitably leads to some polydispersity that decreases at higher
rotation speed as shown on Fig.~\ref{fig2}.

Fig.~\ref{fig:inverted} shows the size distributions of the sphere radii for
several rotation speeds~$\omega$. The emulsions were prepared with
$R_\mathrm{mixer}=1.6\:\mathrm{cm}$, $\phi=0.8$ and $\eta_c=1\mPas$ (same
experiments as the pictures and the red circles in Fig.\ref{fig3}).
Different rotation speeds lead to different distributions, but they collapse to
a single curve when the sizes are rescaled by their average. This gives some
hope that the distribution can be used to discriminate between the different
theories.

We find that the data are slightly better fitted by a log-normal distribution
than by a gamma distribution: The error in the fits, quantified as the summed
square errors between data and model, behave as~$1\div 1.6$ (or as~$1\div 2.3$
when fits were done to cumulative density functions instead of density
functions). None of the distributions looks like a power-law behavior. This
trend, namely that log-normal fits slightly better than gamma, was equally
present when we fitted individual curves (not shown). Both fitting functions
appear to underestimate the steep ascent at small values of the radius; this is
perhaps due to the finite resolution of the microscope.

Within the range of fluctuations we can say that the curves in
Fig.~\ref{fig:inverted} are all the same. In other words, the distribution,
which is quantified by one scale and maybe several shape parameters, here all
have the same shape parameters. It appears therefore as a valid approximation
that the effect of rotation frequency on the emulsification is governed only by
the average drop size and not by the full shape of its distribution. We found
similar observations also for the influence of viscosity, volume fraction, and
mixer size (not shown).

\subsection*{Variation of the mixer radius}

The emulsion is subsequently prepared with a smaller stator of radius
$R_\mathrm{mixer}=0.8\cm$. As previously, the emulsion is examined at
successive rotation speeds, and the mean droplet diameters are shown as blue
dots in Fig.~\ref{fig3}. Also here the mean droplet diameter scales as the
inverse of the rotation speed. The figure further shows that for a given
rotation speed the smaller mixer radius produces larger droplets.

As a first guess, one could think that breakup occurs when inertial forces are
balanced by viscous forces. The ratio of these forces corresponds to the
Reynolds number $\Reynolds = \rho_c \omega R_\mathrm{mixer} \langle d \rangle /
\eta_c$, where $\rho_c$~is the continuous phase density and $\omega
R_\mathrm{mixer}$~the speed at the tip of the blades. Thus, the balance between
inertial and viscous forces leads to $\langle d \rangle \, \sim  \, \eta_c /
\rho \omega R_\mathrm{mixer}$. This scaling is coherent with the experiments as
the droplet diameter decreases with the rotation speed and the mixer radius. 

\subsection*{Impact of the continuous phase viscosity}

To further evaluate the previous scaling, we varied the viscosity $\eta_c$ of
the continuous phase. To this end, we dissolved $3\wtpc$ of SDS either in a
water solution ($\eta_c=1\mPas$ or in a water/glycerol mixture ($\eta_c=6$ or
$11\mPas$ with $50$~or $60\wtpc$ of glycerol respectively). The emulsions were
then prepared with a mixer of radius $R_\mathrm{mixer}=0.8\cm$.
\begin{figure}
    \centering
    \includegraphics[height=12cm]{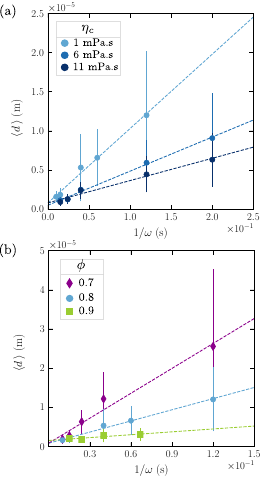}
    \caption{(a) By means of glycerol, the emulsions are prepared with varying
    continuous phase viscosity $\eta_c=1$, $6$ and $11\mPas$ (from light to
    dark blue). The droplet size decreases with $\eta_c$. (b) Mean droplet
    diameter $\langle d \rangle$ as a function of the inverse of the rotation
    speed $\omega$ for volume fraction $\phi=0.7$ (purple diamonds),
    $0.8$~(blue dots) and $0.9$~(green squares). The emulsions are prepared
    with a mixer of radius $R_\mathrm{mixer}=0.8\cm$ and a continuous phase of
    viscosity $\eta_c=1\mPas$. Again, vertical bars indicate the standard
    deviation.}
    \label{fig4}
\end{figure}

For each continuous phase viscosity, we measured the mean droplet diameter at
successive rotation speeds. Fig.~\ref{fig4}(a) shows the mean droplet diameter
as a function of the inverse of the rotation speed for $\eta_c=1$, $6$ and
$11\mPas$ (from light to dark blue). For a given rotation speed, the droplet
size decreases with the viscosity. Hence, higher viscous forces result in
smaller droplets. This trend contradicts our initial guess that the Reynolds
number might be the relevant scale in the droplet breakup.

As an alternative hypothesis for the breakup, we now concentrate on the surface
tension forces, compared to viscous stresses. The viscous stresses can either
be modulated by the viscosity of the continuous phase, or as an effective
viscosity varied by different volume fractions for the dispersed phase.

\subsection*{Variation of the Volume fraction}
To investigate the impact of the volume fraction we disperse $70$, $80$ or
$90\wtpc$ of oil in the continuous phase. As previously, the total amount of
oil is added to the aqueous phase at low rotation speed. The latter is then
increased while samples of the emulsion are collected. Fig.~\ref{fig4}(b) shows
the mean droplet diameter as a function of the inverse of the rotation speed
for varying volume fraction $\phi$. The emulsions are prepared with a mixer of
radius $R_\mathrm{mixer}=0.8\cm$ and a continuous phase of viscosity
$\eta_c=1\mPas$.

For a given rotation speed the droplet size decreases with the volume fraction. Indeed, if a larger volume of oil 
is dispersed, the effective viscosity is increased. Thus, higher viscous forces are exerted and the droplets are smaller.

As viscous forces have a great impact on the size, one can wonder if the droplet breakup could result from a local mechanism governed by a balance between viscous and capillary forces. 

\subsection*{Capillary number scaling}
The process that breaks larger droplets into smaller droplets has to act
against the surface tension. The stress to deform a droplet of size~$d$, is
$\gamma\kappa\propto\gamma/d$, where $\kappa$~is its curvature. If the
mechanism of droplet breakup comes from viscous shear stresses, which are
largest when the emulsion passes through the stirrer's gap of size~$L$, then
these stresses scale as $|\eta\nabla\textbf{v}|\propto \eta\omega
R_\text{mixer}/L$. Their ratio forms the capillary number
\begin{equation*}
    \text{Ca} = \frac{\eta\omega R_\text{mixer}\langle d\rangle}{\gamma L},
\end{equation*}
and droplets break in the given shear stress as long as $\text{Ca}>1$, until
they are small enough such that the surface tension can keep them in shape. We
therefore expect the average droplet size to scale as
\begin{equation}
  \label{Ca_scaling}
  \langle d\rangle \sim \text{Ca}\,\frac{\gamma L}{\eta\omega R_\text{mixer}}.
\end{equation}

\subsubsection*{Interfacial tension measurements}

Measuring the interfacial tension between the oil and the continuous phase is not straightforward, since it depends on the concentration of the surfactant~SDS which changes during the emulsification process. The surfactants initially present in the continuous phase accumulate on the surfaces of the droplets, thereby establishing a dynamical equilibrium that relates the SDS~concentrations on the surfaces to that in the bulk.

\begin{figure}[h!]
    \centering
    \includegraphics[height=11cm]{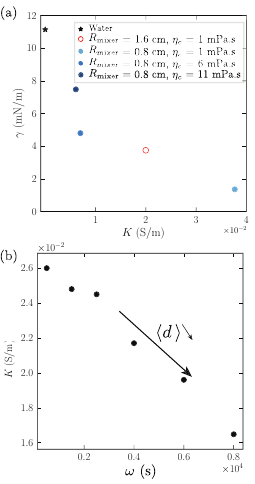}
    \caption{(a) Interfacial tension $\gamma$ as a function of the conductivity of the emulsion. The tension is measured with the inverted pendant drop method using an oil drop immersed in a water/SDS solution mimicking the continuous phase of the emulsion. (b) The conductivity~$K$ of the emulsion as a function of rotation speeds. The emulsion has volume fraction~$\phi=0.65$, with $1\wtpc$ of SDS, generated in a mixer of radius~$0.8\:\text{cm}$.}
    \label{fig5}
\end{figure}

The sodium ions from the SDS make the solution conducting, resulting in a monotonically increasing relation between them. We can therefore conclude on the bulk SDS~concentration by measuring the conductivity of the emulsion.
Vice-versa, if we happen to measure the same conductivity in a homogeneous solution of SDS as in a given emulsion, then we know they have the same bulk concentrations of SDS. This trick allows us to determine the interfacial tension of the droplets in the emulsion: We measure instead the tension of a macroscopic oil drop, deformed under the effect of gravity, by the so-called ``inverted pendant drop method'' in a homogeneous environment of SDS whose conductivity matches that of the emulsion. The result is shown in Fig.~\ref{fig5}(a). As expected, the tension decreases with increasing conductivity from $\gamma=11\,\text{mN}/\text{m}$ for $K=0.05\,\upmu\text{S}/\text{cm}$ (pure water) to $\gamma=1.4\,\text{mN}/\text{m}$ for $K=377\,\upmu\text{S}/\text{cm}$.
As the rotation speed increases, the droplets become smaller thus more and more surfactants leave the bulk to cover the interfaces, resulting in a decrease of the emulsion conductivity. Figure~\ref{fig5}(b) shows this effect: the conductivity~$K$, being directly related to the SDS~concentration, is found to decrease as a function of the rotation speed~$\omega$. 

\subsubsection*{Scaling}
\begin{figure}[h!]
    \centering
    \includegraphics[height=5.5cm]{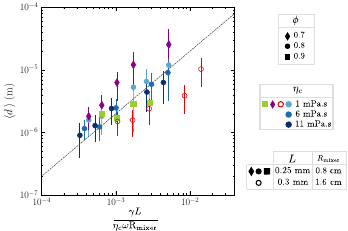}
    \caption{Rescaling of the droplet sizes in terms of the capillary number.
    The liquid parameters vary according to the legend. The dashed line is a
    guide to the eye to show linear scaling. Vertical offsets from the dashed
    line mean that the capillary number in Eq.~\eqref{Ca_scaling} is not unity.}
    \label{fig6}
\end{figure}

Using the above measurements of the surface tension, we can now check
the role of the capillary number in the droplet creation. In Fig.~\ref{fig6} we
plot the mean droplet diameter against the length scale given by the capillary
number and the other parameters, Eq.~\eqref{Ca_scaling}.

We observe that the linear scaling is found as expected, but it does
not occur precisely at $\text{Ca}\sim 1$. Instead, the breakup is shifted to
higher capillary numbers for less concentrated suspensions, and it is shifted
to lower capillary numbers for larger mixer radii. At the moment we have no
simple scaling argument for these deviations, we report them simply as an
observation.

\section*{Conclusion}

The droplet size distribution of dense castor oil-in-water emulsions stabilized
by SDS surfactant was studied in the absence of droplet coalescence. It was
found that the droplet size greatly varies with the rotation speed, the mixer
geometry and the continuous phase viscosity. The droplet size distribution
was found to depend mainly on the average droplet size, not so much
changing its shape. Furthermore, it is somewhat more accurately described by a
log-normal distribution rather than by a Gamma distribution; the former is
expected for the breakup of droplet in a turbulent flow, the latter suggest the
formation of ligaments that subsequently break up in the flow. It would
therefore be interesting to look into the details of the drop formation
process, e.g. with numerical simulations \cite{toschi}. A simple scaling for
the mean droplet diameter was derived without making strong assumptions on the
complex turbulent flow. The mean droplet diameter was found to scale
linearly in the lengthscale that relates surface tension to shear stresses. The
precise value of the capillary number, however, still depends on the other
material parameters of the emulsion. As such mixers are commonly used for
emulsion manufacturing, the results here should be useful for controlling the
rheology of emulsions for food, cosmetics and various industrial applications.
In addition, their stability, adhesive properties \cite{AMAR20051} and
elasticity \cite{Dinkgreve2017}, are also controlled by the drop size and
volume fraction.\\

\bibliographystyle{ieeetr}
\bibliography{bibliography}
\end{document}